\documentclass[Afour,sageh,times]{sagej}

\usepackage{moreverb,url}

\usepackage[colorlinks,bookmarksopen,bookmarksnumbered,citecolor=black,urlcolor=black]{hyperref}

\newcommand\BibTeX{{\rmfamily B\kern-.05em \textsc{i\kern-.025em b}\kern-.08em
T\kern-.1667em\lower.7ex\hbox{E}\kern-.125emX}}

\usepackage{balance}       
\usepackage{graphics}      
\usepackage[T1]{fontenc}   
\usepackage{txfonts}
\usepackage{mathptmx}
\usepackage{tabularx}
\usepackage{color}
\usepackage{colortbl}
\definecolor{lightgray}{gray}{0.9}
\usepackage{array}
\newcolumntype{L}[1]{>{\raggedright\let\newline\\\arraybackslash\hspace{0pt}}m{#1}}
\newcolumntype{C}[1]{>{\centering\let\newline\\\arraybackslash\hspace{0pt}}m{#1}}
\newcolumntype{R}[1]{>{\raggedleft\let\newline\\\arraybackslash\hspace{0pt}}m{#1}}
\newcolumntype{B}{@{\extracolsep{0.5cm}}c@{\extracolsep{0pt}}}%

\usepackage{booktabs}
\usepackage{textcomp}

\usepackage{multirow}

\usepackage{microtype}        
\usepackage{ccicons}          
\usepackage{lscape}
\usepackage{graphicx}

\usepackage[normalem]{ulem}

\newcommand\clearrow{\global\let\rowmac\relax}
\clearrow

\definecolor{linkColor}{RGB}{6,125,233}

\def\volumeyear{2016}

\begin{document}
\newcommand{\subsubsubsection}[1]{\textbf{#1.}}

\runninghead{Anonymous Authors}

\title{Online Knowledge Production in Polarized Political Memes: The Case of Critical Race Theory}

\author{Alyvia Walters, Tawfiq Ammari, Kiran Garimella, and Shagun Jhaver\affilnum{1}}

\affiliation{\affilnum{1}Rutgers University, USA
}

\begin{abstract}
Visual culture has long been deployed by actors across the political spectrum as tools of political mobilization, and have recently incorporated new communication tools, such as memes, GIFs, and emojis. In this study, we analyze the top-circulated Facebook memes relating to critical race theory (CRT) from May 2021 -- May 2022 to investigate their visual and textual appeals. Using image clustering techniques and critical discourse analysis, we find that both pro- and anti-CRT memes deploy similar rhetorical tactics to make bifurcating arguments, most of which do not pertain to the academic formulations of CRT. Instead, these memes manipulate definitions of racism and antiracism to appeal to their respective audiences. We argue that labeling such discursive practices as simply a symptom of ``post-truth'' politics is a potentially unproductive stance. Instead, theorizing the knowledge-building practices of these memes through a lens of political epistemology allows us to understand how they produce meaning. 
\end{abstract}


\maketitle

\section{Introduction}

Critical race theory, a once-esoteric legal theory, came to public consciousness at least partially through the 2021 Virginia gubernatorial campaign \citep{Barakat_Rankin_2022}. Republican hopeful Glenn Youngkin vowed to ban CRT from public education on his first day in office, a promise upon which he delivered with an executive order after his election. Though there is not strong enough evidence to say whether campaigning on critical race theory is what actually won Youngkin the election, there is no doubt that it was a major platform stance that picked up traction both in the state and across the nation \citep{Beauchamp_2021b}. From this gubernatorial campaign, a flurry of public debates surrounding CRT have ensued in political speech, policy proposals, and social media discourse.


Digital social media platforms have been a hub for this discussion where users often share their opinions, beliefs, and concerns about CRT through memes. For this study, we collected 5,662 CRT-centered memes, which circulated on Facebook from May 2021 -- May 2022, and qualitatively analyzed the 27 top-circulated memes. We aimed to gain insights into the discourse surrounding CRT and examine how visual means are deployed to influence the public by analyzing the images shared by users. 
Scrutinizing the dominant themes and narratives in these images, we found that CRT memes provided a site of overt epistemological struggle: both pro- and anti-CRT camps used similar rhetorical tools to make bifurcating arguments about the stakes of integrating “CRT” into the American socio-cultural landscape. While doing so, these ideological groups competed for the most convincing definitions of both critical race theory and antiracism which a) fit their own political framings, and b) worked to convince their community of followers that their beliefs were not-racist. As such, the institutional definitions of both critical race theory and antiracism were largely absent from these memes.
We argue that it is erroneous to mark memes that get these definitions ``wrong'' as simple artifacts of a post-truth society. Instead, we advocate for a more critical look at the ramifications of such visual cultural artifacts through the lens of political epistemology. We posit that these memes are rhetorically complex units of sensegiving that are performing significant political work for both supporters and opponents of CRT.


\section{Background and Related Work}

While Governor Youngkin may have brought critical race theory into public consciousness, it is evident that the socio-legal conceptualizations of CRT are not the same as those that are causing political divisiveness in our current climate. As such, our analysis is dependent on understanding a) what CRT is, really, b) how widely-circulating public discourse on CRT diverges from these established definitions of it, and c) how these public definitions are demarcating the bounds of political in-groups and out-groups in memes.

Critical discourse on race and social media is well-developed, and researchers in the field have investigated a range of pertinent topics, including how conversations on race and racism circulate online \citep{Carney_2016, Matamoros-Fernández_2017, Moody-Ramirez_Tait_Bland_2021} as well as the interpersonal \citep{Cestone_Jones_Harris_Quezada_Roest-Gyimah_2022, Lee-Won_White_Potocki_2017} and social effects \citep{Ray_Brown_Fraistat_Summers_2017, Noble_2018} of these discourses. The current study aims to contribute to this literature by analyzing the rhetorical tools through which critical race theory was defined and circulated in Facebook memes, and ends in a discussion of the social significance of this process.

\subsection{What is Critical Race Theory?}
Critical race theory (CRT) was established in the 1970s when a group of lawyers, activists, and legal scholars began questioning why the constitutional victories of the civil rights era were stalling, or even seemingly being disintegrated \citep{Delgado_Stefancic_2023}. In response to these concerns, CRT posits that the legal system, specifically, but political institutions at large are designed to support whites while marginalizing non-whites in both obvious and coded ways. As Cornel West defines it, CRT is ``the historical centrality and complicity of law in upholding white supremacy (and concomitant hierarchies of gender, class, and sexual orientation)'' \citep[p. xii]{West_1996}. Taking law as a political agent rather than a neutral power structure, critical race theorists investigate how social institutions create and uphold racism, and with a strong activist dimension, they also seek to change these conditions.

As will soon be evident, this legal studies definition is not materializing within most memes considered in this study. However, it does not do to simply state that these memes are incorrect. While they are technically incorrect in an institutional sense, they still make meaning for their audiences--and this community sense of CRT, which gained visibility via artifacts such as memes, may be more politically relevant than the lesser-known institutional definition. 

\subsection{How Communities Make Sense of Things: Knowledge-Building and Epistemology}
Communities build knowledge through a shared understanding of the world and often a shared value system. However, in the current U.S. political climate, this shared knowledge-building is often not based on credible fact, which has led scholars to develop notions of ``post-truth'' societies and ``fake news'' \citep{Rose_2017}. Ways of knowing, or epistemologies, are one lens through which we can discuss the construction of community-built knowledge, and in this case, grassroots understandings of critical race theory. 

``Political epistemology'' is a growing area of research that brings together scholars who are interested in the intersections of political philosophy and epistemology. This juncture provides fertile space to investigate topics such as misinformation, polarization, and the ``epistemic virtues (and vices) of citizens, politicians, and political institutions'' \citep[p.1]{Edenberg_Hannon_2021}. The moment we are analyzing --- one in which critical race theory is being politicized --- lends itself well to theorizations of how political ``ways of knowing'' materialize and what stakes these epistemologies may have.

Of growing interest in studies of political discourse is what is referred to as the ``post-truth'' age. The conception of post-truth is directly tied to conceptions of political epistemology because many scholars argue that ways of knowing have been complicated by rising disregard, disbelief, or lack of interest in truth \citep{McIntyre_2018}. In a related vein, ``bullshit'' has also been theorized as a contemporary way of doing politics and can range, discursively, from rambling on about topics that one knows nothing about to crafting complex lies with specific end goals in mind (Cohen, 2002; Frankfurt, 2005; Lackey, 2007). 

However, \cite{Cassam_2021} argues that the very ideas of ``post-truth'' or ``bullshit'' as tools of political epistemology hold far less weight than others suggest and likely do justice to neither the complicated rhetoric deployed by politicians nor to the public's reaction to these techniques. He questions their effectiveness as tools of description or explanation in political discourse, and he argues that what is usually described as post-truth or bullshit is often far better captured through the lenses of hate speech or propaganda analysis. He writes, ``It is a travesty to describe hate speech as mere bullshit since this does not even come close to capturing what is wrong with it and why it works.'' \cite{Cassam_2021} is not arguing that post-truth and bullshit are not useful concepts, but rather that it is a grave mistake to subsume political epistemological analysis --- particularly post-2016, when much discussion of the post-truth politics came to the fore --- under the assumed post-truth umbrella. For the purposes of this study, we extend this notion to not only politicians but to those who are disseminating politicized information, as well. We question how useful it is to write off the mis-/disinformation provided in the memes under study as yet another manifestation of post-truth politics.

\subsection{Political Memes as Objects of Sensemaking}
Memes, as defined by Limor Shifman, are ``units of popular culture that are circulated, imitated, and transformed by individual Internet users, creating a shared cultural experience in the process'' \citep[p. 367]{Shifman_2013}. Due to the grassroots nature of memes, internet circulation of meme-based information stands in stark contrast to that of ``media elites:'' a status which may lend a level of authenticity not otherwise afforded to traditional media discourse (Burroughs, 2020). Political memes often work to make complex arguments more digestible for a broad audience. They are thus valuable to study for their ability to ``[connect] the political to the popular, the political to emotionally charged, affective media'' \citep[p. 192]{Burroughs_2020}.

In what \cite{Lankshear_Knobel_2019} deem the ``second wave'' of online memes, the use of memes as political sensemaking tools, which are often weapons in sociocultural wars, looms large. In recent works, \citet{Ross_Rivers_2018} found that political memes reflected in-group tensions throughout the 2016 US presidential primaries, and were subsequently used to delegitimize both candidates--thus creating lines of in- and out-group online communities--in the general campaign cycle. \citet{Woods_Hahner_2020} analyzed how the alt-right uses memes to continually re-make what is deemed acceptable discourse on the political right, thus lending authority and sense to increasingly extreme rhetorics within the group bounds. In a study closely related to our own, \cite{Moody-Ramirez_Tait_Bland_2021} investigated memes featuring  information on race, oppression, and protest following the marches in response to George Floyd's death in the summer of 2020. They found that memes were a site in which competing senses of reality were being constructed, with often-racist framings of these protests coming to the fore.

As objects of sensemaking, the question of why memes are ripe sites for deepening political divides is pertinent. According to \cite{Dean_2018}, memes have the potential to serve an Althusserian \textit{interpellative} function. By this, he means they `` `hail' the viewer into identifying with  them, either by agreeing with the political sentiments expressed therein, or by finding them funny (or not).'' He argues that memes can consolidate political allegiance, entrench antagonisms, and shape political discourse due to their punchy, shareable nature. \cite{Askanius_2021} agrees, noting that the visual aspect of memes makes them highly transmissible because images have the capacity to cut across cultural and linguistic barriers. This easy access can ``foster a sense of community and belonging...allow[ing] a target audience to be `in' on the joke and self-identify with the message of that meme.'' (p. 116) In the case of fringe ideologies, this sense of belonging can serve as a ``gateway'' to deeper radicalization and divide \citep{Askanius_2021} through similar appeals that analog, leaflet propaganda made: promises and affirmations of users' sense of tribalism \citep{Nieubuurt_2021}.

The current study contributes to this literature through its attention to the sensemaking functions of memes following the political eruption of CRT in the early 2020s. Through a mixed-methods approach, we claim that these highly-circulated CRT memes compete for validity by using parallel rhetorical tools to define what CRT is, but ultimately land on vastly different definitions in order to accrue in-group approval and make sense of this political flashpoint.
\section{Methods}

\subsection{Data Collection}
In this work, we focused on the popular images shared in the discussion around critical race theory. To identify these images, we collected public Facebook posts and images published between May 2021 and May 2022 which were discussing critical race theory. During that year, there had been multiple spikes in the discussion around CRT on Facebook in line with real-world events like the Virginia Gubernatorial Election, making this time span appropriate for analysis.

We used \citet{crowdtangle}, a tool provided by Meta that enables searching and analyzing public content from Facebook.
We collected all posts from Facebook that contained the term “critical race theory”, and had a minimum of 100 interactions, as we were interested in analyzing the images with the largest reach. 
We did not include the term “CRT”, a popular abbreviation of critical race theory, in our search query as our early sampling and search results review indicated a high false positive rate for that term (e.g. related to CRT televisions). 
This gave us 5,662 posts during the period May 2021-May 2022.
Since a majority of the posts (around 70\%) contained images, we decided to focus on images.
The final dataset consisted of 3,906 images that were accessible and downloadable.\footnote{We will share a link to this dataset after the peer review is completed.} 

\subsection{Clustering}
Once all the images were collected, the next step was to identify the popular images among them. We defined an image's popularity as the number of times an image appears in our dataset. We borrowed \citet{zannettou2018origins}’s method of using image hashing, specifically pHash~\citep{1709989} values to identify similar images. pHash is an algorithm for perceptual hashing~\citep{farid2021overview} which returns a random string (`hash') for any given image. The property of this random string is that perceptually similar images (e.g. images that are slightly cropped, or have a watermark but are otherwise the same image) have similar pHash values.
Given the pHash values for two images, we can compute the distance between them to infer if the two images are similar. 

Clustering is a technique to identify and group similar objects based on a specific property into the same cluster. We used DBSCAN~\citep{10.5555/3001460.3001507}, a density-based clustering algorithm to group the identical images. DBSCAN considers clusters to be dense regions of data points, handles well the clusters of arbitrary shapes and is also robust to noise and outliers. We performed clustering based on the distance between the hashes, which gave us 190 clusters. Each cluster had multiple images in it, with the cluster size ranging from 3 images to 28 images.


\subsection{Coding \& Critical Discourse Analysis}
\subsubsection{Qualitative Content Analysis}
Because we undertook an iterative image coding process, we included enough images in the analysis to reach thematic saturation \citep{Low_2019}, which was 35 clusters. Within this set of 35, several clusters/images were so rhetorically similar that we collapsed them into one category, leaving us with 27 distinct images for analysis. Initially, images were considered apart from their contextualizing captions and comments for analysis, but in cases where it was not particularly clear which code an image should be given, we considered the surrounding text and reactions on the Facebook post where the image was shared to get a better understanding.
 
We carried out the analysis in an iterative manner. First, we decided on the categories/dimensions for which the images should be coded, the most basic being a binary categorization of pro-CRT or anti-CRT. Then, we began qualitatively coding for emergent themes. These codes were refined over multiple iterations until we finally grouped similar codes together to create organized parent codes.  Though we created parent codes for multiple image categories (e.g., `type', `origin'), we primarily focused on the ``role'' of images.
The set of ``role'' codes captures how the image is deployed and the message/intent the image is attempting to convey.

\subsubsection{Critical Discourse Analysis}
After coding these memes to better understand their rhetorical functions, we finally engaged critical discourse analysis (CDA) in order to introduce questions of power in our semiosis. In this methodology, language is never read as neutral and is instead analyzed for its ideological underpinnings. According to Fairclough, CDA
\begin{quote}
provides a methodology to systematically explore often opaque relationships of causality and determination between (a) discursive practices, events and texts, and (b) wider social and cultural structures, relations and processes; to investigate how such practices, events and texts arise out of and are ideologically shaped by relations of power and struggles over power . \citet [p. 93] {Fairclough_2018}
\end{quote}

He theorizes discourse as a “three-dimensional” structure which is made up of discourse events, discursive practices, and social practice. Discourse events, he posits, are the actual “text” to be analyzed—“text” meaning any culturally-situated object of study—and these discourse events are composed of both discursive practices and social practices \citep{Fairclough_2018}. 

In the case of the present study, memes are the discourse events which we analyze for their discursive practices--what the text and image, together, are discursively creating and reflecting--and for their social practices--how these discourses are tied up in sociocultural contexts. The content analysis allowed us to see trends in the discursive practices of these memes, and situating these trends within the social contexts of political and hegemonic power relations allowed us to make our ultimate arguments on why these memes, as discourse moments, matter in a crowded field of political discourse.

\section{The Tools of CRT Meme Production}

After iteratively coding each meme, we were left with 21 unique codes which could be subsumed under one or more of the following major rhetorical tactics: 1) struggles over definition, or how the meme makes sense of what critical race theory is; 2) constructing “antiracism,” or the ways in which the meme’s ideologies are coded as definitively not-racist to its intended audience; and 3) appeals to authority, or the ways in which the meme uses people or symbols to appear correct. These strategies, then, appear to be the most salient paths through which both pro- and anti-CRT arguments within these memes are built.

\subsection{Defining CRT}
With two exceptions, neither the pro- nor anti-CRT memes analyzed appear to be concerned with the “real” critical legal studies definition of critical race theory. 
%
Because CRT was not generally circulated within public discourse prior to the early 2020s, as mentioned above, there was a wide berth for political and epistemological work to be done in the construction of this definition in the public consciousness. Our analysis reveals that these memes do just that: while both pro- and anti-CRT memes provide a technically incorrect definition of what CRT actually is, the politics of sensemaking unfolds within these memes. Their consumers are left with bifurcating definitions of the bounds, risks, and benefits of critical race theory that ultimately serve to re-define the bounds and values of the communities in which these memes circulate.
    
\begin{figure}[htp]
    \centering
    \includegraphics[width=0.5\linewidth]{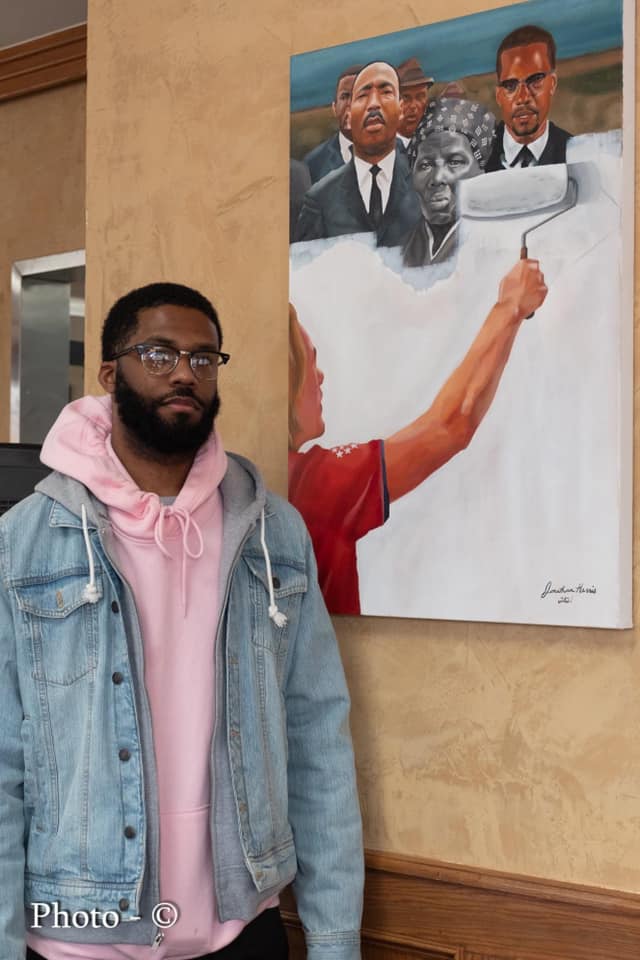}
   \caption{Pro-CRT meme of artist Jonathan Harris with his painting entitled "Critical Race Theory"}
    \label{fig:cluster1}
\end{figure}

Within pro-CRT memes, the aggregated definition reads something like this: \textit{critical race theory means teaching history accurately and not being a racist and/or a Republican.} 
These memes went to far fewer lengths than anti-CRT memes to define what CRT actually is, and relied instead upon defining it against other things: racism, Republicans, and/or the erasure of history. The top-circulated meme in our analysis is a prime example of this. In this meme, artist Jonathan Harris stands alongside his now-viral artwork entitled “Critical Race Theory,” which depicts the literal whitewashing of Black history (Figure \ref{fig:cluster1}).

\begin{figure}[htp]
    \centering
    \includegraphics[width=0.5\linewidth]{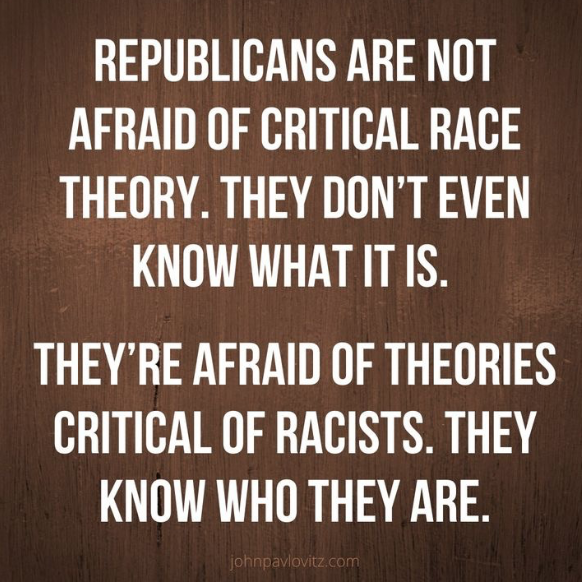}
   \caption{Pro-CRT meme accusing Republicans of being racist and ignorant}
    \label{fig:cluster4}
\end{figure}

While this does nothing to define critical race theory, it certainly defines what it is not: the erasure of America’s violent, racist past. Similarly, image cluster 4 (Figure \ref{fig:cluster4}) is a computer-generated text-heavy meme that reads, “Republicans are not afraid of critical race theory. They don’t even know what it is. They’re afraid of theories critical of racists. They know who they are.” The irony, of course, is that this meme also does not indicate a real definition of CRT, or an indication of “knowing what it is”--it simply defines CRT against racists and Republicans, both of which believers in CRT cannot be. 

In contrast to this strategy, anti-CRT memes often utilize quite specific points of definition. Take, for instance, cluster 22 (Figure \ref{fig:cluster22}): 

\begin{figure}[htp]
    \centering
    \includegraphics[width=0.5\linewidth]{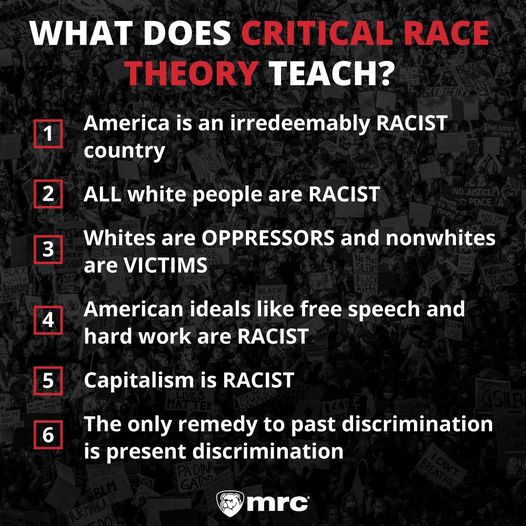}
    \caption{Anti-CRT meme outlining CRT's alleged values}
    \label{fig:cluster22}
\end{figure}

\begin{figure}
    \centering
    \includegraphics[width=0.5\linewidth]{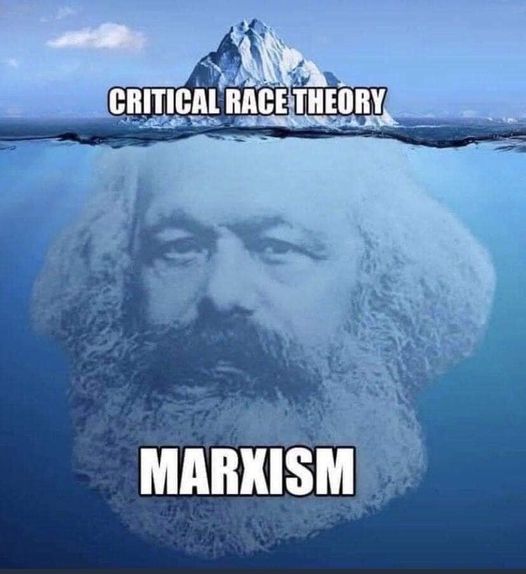}
    \caption{Anti-CRT meme connecting CRT to Marxism}
    \label{cluster16}
\end{figure}

\begin{figure}
    \centering
    \includegraphics[width=0.5\linewidth]{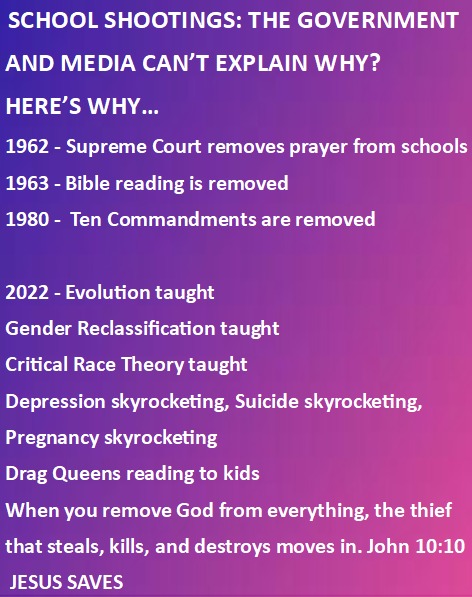}
    \caption{Anti-CRT meme suggesting a connection between ``wokeness'' and school shootings}
    \label{cluster20}
\end{figure}

In providing a six-point bulleted list, this meme lays out, in no uncertain terms, how its consumers are meant to understand CRT. These definitional strategies, however, are not always so concrete as a bulleted list. Whereas pro-CRT memes defined CRT \textit{against} other ideas, anti-CRT memes sometimes worked to define it by conflating it \textit{with} other ``anti-American'' ideas, such as Marxism (Figure \ref{cluster16})  and straying from Christianity (Figure \ref{cluster20}). In other words, anti-CRT memes often tied CRT to other ``woke'' ideologies in order to define it, even as ``wokeness,'' in and of itself, lacks definitive boundaries. 

In all, anti-CRT memes essentially define critical race theory in the following way: \textit{CRT is a racist idea that makes people believe that race matters more than it should, and it is yet another way that ``wokeness'' is destroying America}. This conceptualization is starkly different than that of the pro-CRT memes, and both are far from the ``real'' legal studies definition, as outlined above. As such, there is obvious political struggle in the fight to win the hegemonic, accepted definition of critical race theory --- a definition which has little to do with its origins in critical theory and law.

\subsubsection{Metadiscourse on the Stakes of Defining CRT}
\begin{figure}
    \centering
    \includegraphics[width=0.5\linewidth]{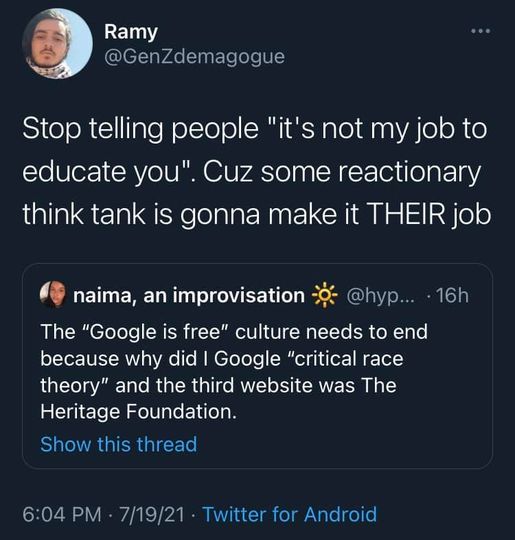}
    \caption{Pro-CRT meme which indicates issues around defining CRT}
    \label{cluster27}
\end{figure}

In this discussion on the political struggle over defining CRT, one particularly interesting meme to highlight is that of Cluster 27 (Figure \ref{cluster27}). 

In this meme, a screenshot of a quoted tweet, two people are explicitly naming this struggle and pointing to its risks. By indicating that The Heritage Foundation, a conservative think tank known for their right-wing ideologies, is a top hit in searching for information on critical race theory,  the original tweet author is pointing out that knowledge acquisition via the internet is deeply politicized, and without careful, critical consumption habits, people can be easily misled by seemingly authoritative information. The quote tweet, agreeing with this view, further interpolates into an ongoing debate in U.S. culture on whose ``job'' it is to educate on topics surrounding race and racism: people of color, who are exhausted by confronting this responsibility every day, but are also the people who have actual experience with racism; or white people, who are the ones who should be expending energy into acquiring knowledge to better educate themselves, without needing to further exploit the time and energy of people of color to do so (see, for example, \cite{Zheng_2021}).

Both Twitter users, ``naima, an improvisation,'' and ``Ramy'' land in the same place: passing this responsibility to educate over to the internet is dangerous territory in a politicized information environment. If ``some reactionary think tank'' such as The Heritage Foundation is where people are gaining their knowledge because ``Google is free'' and no one else is providing this information, the struggle over defining these words--words which have actual policy impact, as seen through Youngkin's Executive Order 1--is of utmost importance, and it appears that memes are one avenue in which this epistemological struggle occurs.

\subsection{Defining ``Antiracism''}
These pro- and anti-CRT battles over definition, and the recognition in the metadiscourse that this is, indeed, a battle, is almost exclusively fought on the same grounds: that of antiracism. While there are some other nods to bigotry in its various forms, for example, transphobia as displayed in Figure~\ref{cluster20}, race and racism are unsurprisingly the main sites upon which definitions of critical race theory and its risks and/or benefits, are built. However, in a similar fashion to how ``critical race theory'' was defined to meet community needs rather than to reflect a ``real'' definition of the term, ``antiracism,'' too, is made into a fungible ideal constructed to meet the dire need of both sides of this argument to appear not-racist: a near necessity in 2023 America.

According to \cite{Ferguson_2022}, antiracism has suffered from a lack of coherent and accessible academic definition. Thus, she proposes a paraphrase of Black author and activist Ijeoma Oluo's tweeted definition: ``the commitment to eradicate racism in all its forms,'' with a noted special interest in recognizing the difference between systemic and interpersonal racism. However, the tension we encounter in the analysis of these images is that, under this definition, both overt racism and a quieter ``not-racism,'' a term we will more clearly define below, is easily able to masquerade as antiracism to the undiscerning eye. In this way, the memes' constructed definitions of antiracism become a technology through which racism is perpetuated.

\cite{Bonilla-Silva_1997}'s oft-cited definition of racism demands recognition of both structural and interpersonal forms of racism, and it also demands acknowledgment of the difference between the two. Because these pro-CRT memes often fail to address structural racism, they fall short of forwarding truly antiracist ideologies even as they consistently present themselves as performing antiracism. These failures come through in what we are deeming ``not-racism,'' or essentially a focus on the interpersonal aspects of racism only, rather than on the structural ones. For instance, in Cluster 4 (Figure \ref{fig:cluster4}), the deflective ``they'' indicates several things all at once: first, that ``we,'' those who identify with this meme, are not like ``them,'' the racist Republicans; and second, that the stakes of this argument on critical race theory reside at the individual, interpersonal level. Each of these two implications constructs racism as something that happens within the hearts and minds of individuals, rather than at the structural level, and further, it absolves those who resonate with the meme from racism: ``I am not a racist, because I support critical race theory.'' This is not-racism in practice: a positionality that is not reflective of true antiracism, which is an ideology and practice that requires recognition of and action toward dismantling systems of racism, interpersonal racism, and the implicit bias that each of us holds. 

This is not to say that pro-CRT memes always failed at performing antiracism, either. There certainly were instances of successful acknowledgement of structural racism, such as in Cluster 7, which reads ``If people attacked White Supremacy like they are attacking critical race theory, there would be no need for critical race theory.'' However, the failures were all failing in the same way: by framing themselves and their ideologies as not-racist, and falsely equating that with doing antiracist work.

On the other hand, anti-CRT memes wholly fail at performing true anti-racism, and they fail in many different ways: through rhetorics of racial neoliberalism, colorblind racism, and not-racism. Importantly, each of these tools of racism are constructed as antiracism, and sold to audiences as such. While not-racism presents a bit differently on this side of the aisle, the main takeaway is meant to be rhetorically the same: ``we'' are not the racists, ``they'' are. For instance, Figure~\ref{fig:cluster22} reads, in part, ``What does critical race theory teach?...The only remedy to past discrimination is present discrimination.'' This swiftly both dismisses racism as in the ``past,'' thus ignoring its structural persistence and constructs those that support critical race theory as the racists.

In this way, both pro- and anti-CRT memes variably fail at performing antiracism, often forwarding what Blake et al. (2019) call ``antiracist appropriation,'' or a strategy that is ``primarily concerned with deciphering who is a racist and who is not, rather than working to dismantle racism's socially shared institutional and affective structures'' (p. 23). By forwarding this claim, we do not mean to engage in an uncritical false balance \citep{Rietdijk2021-RIEPFB} analysis here, as there is clearly one group that is getting closer to actual antiracism than the other: pro-CRT memes. It is important to note, however, that even pro-CRT memes are not actually accomplishing an antiracist agenda, as they are rather uncritical of the structural aspects of racism and choose to focus, instead, on interpersonal-level issues. 

\subsubsection{Appeals to Black Authority}
Though using appeals to authority is not a groundbreaking rhetorical strategy and is, in fact, one of the pillars of Aristotelian rhetorical philosophy, the ways in which this ethos appears within these memes present an interesting finding: equally often, both pro-and anti-CRT memes deployed the imagery and/or quotes of Black people. Through circulating these images widely, those captured in these memes essentially stand in as Black spokespeople for each side of the argument, lending credence to the meme’s ideology–no matter the side of the argument–through the color of their skin. 

\begin{figure}
    \centering
    \includegraphics[width=0.5\linewidth]{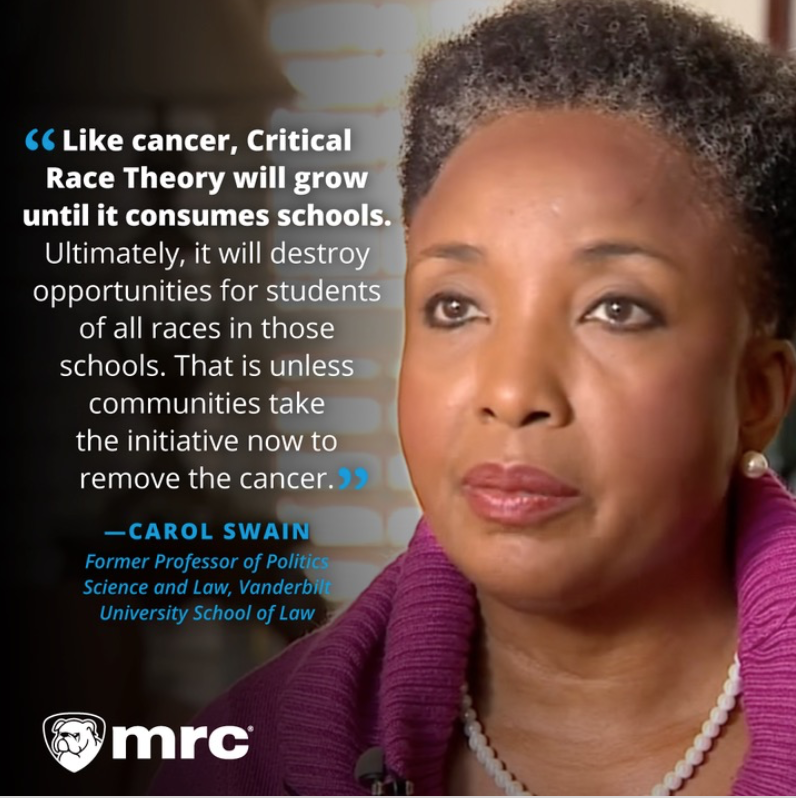}
    \caption{Anti-CRT meme which depicts Professor Carol Swain as a Black woman critical of CRT}
    \label{cluster9}
\end{figure}

Anti-CRT memes that used this rhetorical strategy–all of which, notably, were produced and originally disseminated by the conservative Media Research Center (MRC)\footnote{The Media Research Center (MRC) is more than simply a Facebook page, which is where many of these other memes are sourced. To the contrary, MRC is an entire conservative media network that self-describes its mission as being in accordance with "America's founding principles and Judeo-Christian values." For more, see mrctv.org }–constructed these Black spokespeople as both authoritative in their experience and authoritative in their Blackness. Alveda King, Civil Rights Leader; Dr. Ben Carson, M.D. and former Secretary of Housing and Urban Development; and Dr. Carol Swain, Ph.D. and professor of political science and law appeared in these memes, each in visage and in quote (see Figure \ref{cluster9} for an example of these memes, each of which followed this aesthetic template). The embodied Black professional positionalities which these people inhabit make it difficult for pro-CRT advocates to argue against their claims---claims which invariably speak to the sure pitfalls of socioculturally adopting critical race theory---and thus a comfortable space of disseminating racism through the rhetorics of not-racism opens up.

Using Black spokespeople to deliver implicitly or explicitly racist information has been a tactic used for decades to make news reporting \citep{Entman_Rojecki_2007}, campaign strategy, and political policy \citep{Mendelberg_2001} appear not-racist.
The implicit suggestion is that if a Black person indicates something is not racist, it must not be. This, then, ``bolsters [whites'] denials that racism still impedes the lives of African Americans'' \cite[p.~106]{Entman_Rojecki_2007} and invites a level of assurance that they, too, are not-racist. In the context of the memes analyzed for this study, these Black spokespeople are consistently reflecting a well-established space of Black conservative thought which taps into individualism, self-help, and egalitarianism as answers to discussions on racism \citep{Lewis_2005}.

Critical race theory--the ``real,'' institutional one--actually warns against this very scenario: CRT argues that constructing Black spokespeople as people who can speak for the entire race is both essentialist and ignorant to the importance of intersectionality \citep{Delgado_Stefancic_2023}. 
While it is true that Black Democrats far outweigh Black Republicans in the electorate \citep{Cox_2022}, and thus pro-CRT memes that deploy Black spokespeople likely reflect a larger share of Black thought, it is still unproductive to count any single person as representative of a race of people. Despite this, however, constructing Black spokespeople through memes--Black spokespeople who are made to appear as \textit{the} reasonable ``Black voice''--was a way of building authority and ``assurances'' for those against CRT that they were not thinking in a racist way.

The pro-CRT memes' authoritative appeals to Blackness operated differently and more diversely. Cluster 1 features Black artist Jonathan Harris with his painting (Figure \ref{fig:cluster1}), Cluster 26 thoughts on the legal system from a Black critical race scholar, and Cluster 23 Black singer/songwriter John Legend's call-to-action. 

\begin{figure}
    \centering
    \includegraphics[width=0.5\linewidth]{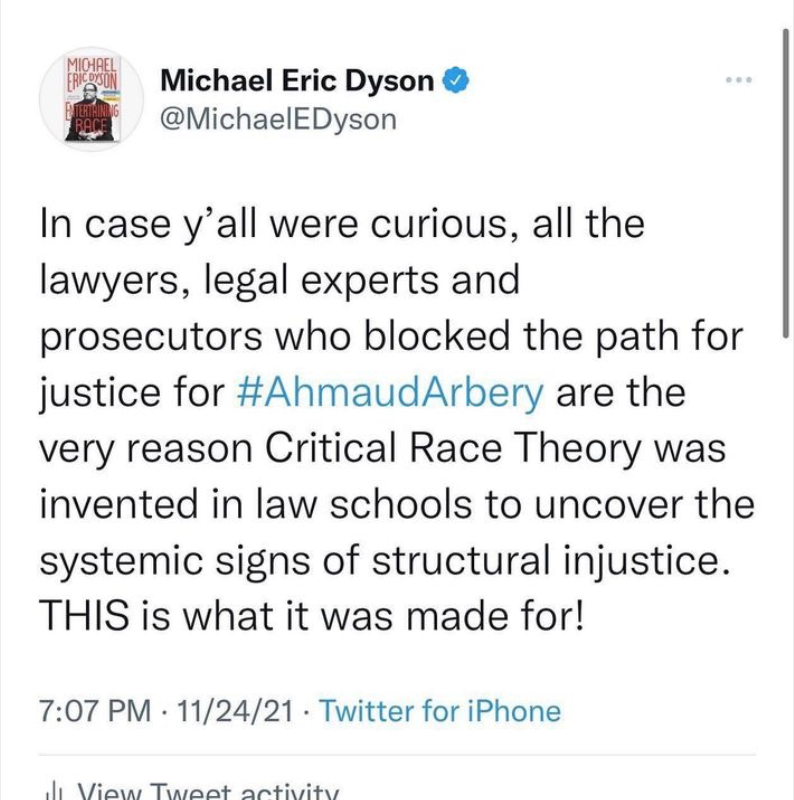}
    \caption{Pro-CRT meme of critical race scholar Michael Eric Dyson's tweet}
    \label{cluster26}
\end{figure}

\begin{figure}
    \centering
    \includegraphics[width=0.5\linewidth]{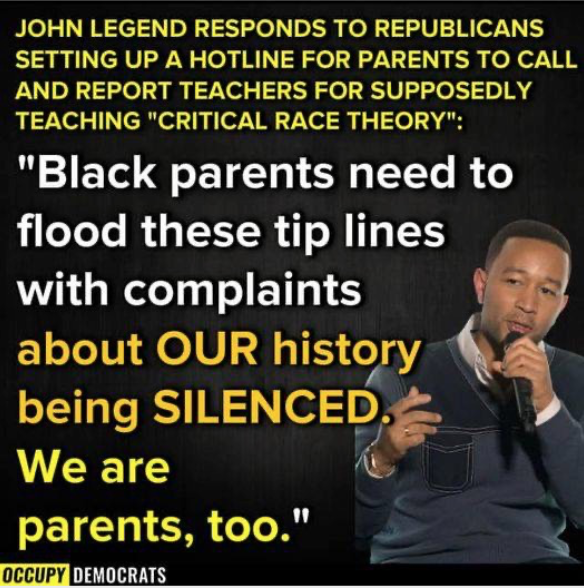}
    \caption{Pro-CRT meme which depicts singer/songwriter John Legend advocating for Black parents to fight back against anti-CRT policy}
    \label{cluster23}
\end{figure}

While their Blackness operates, rhetorically, as an appeal to authority in a conversation about race, the idea of the ``Black spokesperson'' takes on a different function and meaning in these pro-CRT memes. Whereas anti-CRT memes were curating quotes from a very specific set of Black spokespeople, seeking out those who have said something condemning CRT, overlaying these quotes on an image of the Black orator, and circulating that image, those on the pro-side of the issue are more often amplifying already-existing media that Black people created. For example, the image of artist Jonathan Harris (Figure \ref{fig:cluster1}) was an organic, pre-existing photograph of the artist posing with his work--not a computer-generated, curated message that was created without his knowledge or consent. Similarly, the meme featuring Michael Eric Dyson's thoughts (Figure \ref{cluster16}) is simply a screenshot of a tweet he chose to write and publish on the internet--again, not something that an outside entity needed to create. The only exception is the meme featuring John Legend (Figure \ref{cluster23}). This meme has a similar aesthetic to the anti-CRT memes in that it is a computer-generated image of Legend alongside a quote about Black parents needing to get involved in conversations around the banning of critical race theory.  This meme represents just one way that anti- and pro-CRT memes perform ``not-racism'' in rhetorically similar ways, even if they have different end goals.

\section{Knowledge Production in a Post-Truth World}

Memes are an excellent vehicle for making hot-button political issues digestible to the average person. However, almost none of the CRT memes we analyzed actually got its definition ``right.'' Instead, it appears that the most salient rhetorical tools across all 27 of these highly-circulated memes revolved around creating the most convincing definitions of both critical race theory, as that was the topic at hand, and antiracism, as convincing others that an opinion is not-racist is the only socially-palatable way to speak on race in the US in the 2020s. That said, what does this lack of attention to institutional definitions mean in a ``post-truth'' society, and what should we do about it?

Political epistemology allows us to theorize on the importance of these memes as sensemaking and sensegiving tools for the public, and allows us to question the importance of community-based political knowledge over ``real'' institutional knowledge. In the context of CRT memes, the institutional definition of critical race theory \textit{does}, of course, matter in a general sense. The work flowing from this definition has produced massive change at institutional and individual levels, and scholars, writers, and activists use these ideas to dismantle oppressive systems globally \citep{Delgado_Stefancic_2023}. However, that definition is not the one doing political work within these memes, and we would be missing the point entirely if we critique them as simple misinformation or fake news. In a fact-checking sense, all of these memes, both anti- and pro-, are largely false, but it is not useful to write them off as such. This technically false information is filling an information void for people who have likely never before heard of critical race theory, and that means it is these definitions--not the institutional ones--that are doing political work. The discourse is not actually about CRT; CRT simply became a catch-all phrase to hold discourse about race. In reality, the discourse revealed in these memes is about how the US should handle race moving forward, and how we define what is racist and what is antiracist. 

The rhetoric within this discussion presents another point of interest: despite research that suggests overt racism has become more acceptable in a post-Trump America, explicit appeals to racism were not present at all, even in anti-CRT images. After the election of Donald Trump as the US President, several studies have indicated that explicit \citep{GanttShafer_2017} or nearly-explicit racism \citep{Schaffner_Macwilliams_Nteta_2018} became a usable mechanism for Republicans in ways that it has not been since the 1950s and 60s \citep{Mendelberg_2001}. These studies argue that the president's rhetoric ushered in a new era of acceptability of overt racism. However, in the case of the highly-circulated CRT memes we analyzed, this does not appear to be true. While we would argue that, based on \cite{Bonilla-Silva_1997}'s definition of racism, anti-CRT memes \textit{are} forwarding a racist agenda, none of those collected use overt racism in the rhetorical style of far-right groups. Instead, they trend toward the more traditionally-palatable implicit rhetorics that have been successful among conservative voters in the past \citep{Mendelberg_2001}. This is, perhaps, a surprising outcome of this analysis, and may indicate that there is still wider-spread conservative appeal toward implicit rather than explicit racism.

While these implicitly racist appeals follow a long tradition of political rhetoric, the context in which they operate has changed. Following the election of Donald Trump and the ushering in of the ``post-truth'' era, we argue that implicit racism, in particular, runs the risk of being classified simply as disinformation. It is disinformation, for example, to posit that ``the only remedy to past discrimination is present discrimination,'' as cluster 22 does, but it is also much more than that: it is a tapping in to white supremacist understandings of what antiracism is. This, according to \cite{Cassam_2021}, is one of the largest risks we run by taking post-truth as a political epistemology. Rather than understanding these rhetorics as an epistemological formulation of conservative politics, post-truth as a lens for understanding our current political moment could misconstrue this type of rhetoric as simple disregard for truth rather than a calculated dog whistle.

This leads us to two main takeaways. First, institutional definitions matter, but they have little material meaning if the public is defining terms otherwise, especially through highly transmissible and easily digestible artifacts such as memes. 
Second, fact-checking as a practice perhaps misses the mark if it only seeks ``truth'' in a traditional sense, thereby framing false claims as dismissible, post-truth politics. While it is important to assert truth in a misinformation landscape, it is perhaps more important to understand what the actual issue at the heart of the political discourse is, what the stakes are, and what the use-value of the term being wielded is in order to disrupt oppressive practices and support emancipatory ones appropriately.

\section{Conclusion}

This study entices us to continue asking how we might confront mis-/disinformation in our current moment. It becomes especially urgent as we encounter the fact that much of the information circulating through highly transmissible media, such as memes, is not only incorrect but is also fungible: in the case of CRT-centered memes, CRT could ``mean'' almost anything race-related to forward each camp's agenda, and seemingly very few care to engage with an institutional definition. When politicized definitions are a practice in power assertion, the discursive work these definitions do--``correct'' or not--is more necessary than ever to understand. As \cite{Cassam_2021} warns us, it is important not to mistake a lack of engagement with ``true'' definitions as a disregard for the truth as a whole, and assuming that mis-/disinformation is merely bullshit is perhaps an unproductive lens through which to view knowledge production. We must take these definitions seriously, as they are, from a political epistemology stance, thoughtfully crafted messages that are ``true'' in some way to their consumers.

There are several fruitful routes that we can identify for further work around the production and consumption of these memes as they relate to knowledge-building practices. In the space of production, contacting those who created these media objects would potentially lend useful insight about how they, as creators, gained their own understanding of CRT, and why they chose to disseminate this information in these particular ways attached to these particular visual formats. The Media Research Center would be an interesting first place to start, as they crafted each of their anti-CRT memes in the same aesthetic format with the same rhetorical appeals to Black spokespeople. Additionally, study of those who consume these memes is warranted to uncover how users' encounters with these media shape their understandings of CRT and their opinions on it. 

Further, we argue that platforms, too, have some responsibility to contextualize memes such as these through content moderation practices. We acknowledge that this is more than a simple technical issue: filters for racist material, for example, would not flag memes as nuanced as these, and indeed, platforms would likely encounter bad publicity around censorship if any of the memes included in this study were removed. However, there are ways to approach this information landscape through socio-technical solutions, such as by providing the public, experts, and other cultural gatekeepers the ability to contextualize information on social networking sites \citep{Morrow_Swire‐Thompson_Polny_Kopec_Wihbey_2022}. By adding ``notes,'' or otherwise interacting with the information in such a way that its complex relationships to institutional fact are immediately evident to users who may encounter that information, platforms could greatly diminish the power of partisan information masquerading as fact. 

Finally, educational curricula and the students who learn from them would deeply benefit from incorporating critical media consumption practices into their core goals and outcomes. It is no longer possible to separate learning from media consumption in the everyday lives of the vast majority of students in the U.S., and we all suffer when there is a lack of commitment to creating critical media consumers who are trained to think before believing--and even more importantly, re-circulating--a politicized meme. Training young people on how mis-/disinformation and hate speech are disguised as fact and/or humor in memes is an important step forward in strengthening our information landscape and democratic future. Teachers are extraordinarily overburdened already, but a curriculum that integrates media literacy as a guiding principle would partially shift the burden from teachers directly and instead task those who guide the direction of school districts nationwide with creating pathways to teach this skill in all subject areas.

There is no simple solution to curtailing the circulation of harmful visual media, as it is neither a purely tech issue nor purely a lack of education: this is a social issue which can only be resolved through the engagement of a wide variety of actors. It is incumbent upon all of us to take these seemingly insignificant memes seriously for their social impact and what they reveal about current ideological trends. It is crucial to better understand how bottom-up knowledge production on politicized topics, such as CRT, occurs on social media, particularly through compact, made-to-share media such as memes. In doing so, we can move beyond a deterministic conception of post-truth politics which generalizes disregard for truth, and instead interrogate the construction of politicized ``truth'' as a sustained process of thoughtful rhetorical decision-making with real-world effects. 

\bibliographystyle{SageH}
\bibliography{refs.bib,moderation_references.bib}

\begin{thebibliography}{41}
\providecommand{\natexlab}[1]{#1}
\providecommand{\url}[1]{\texttt{#1}}
\providecommand{\urlprefix}{URL }
\expandafter\ifx\csname urlstyle\endcsname\relax
  \providecommand{\doi}[1]{DOI:\discretionary{}{}{}#1}\else
  \providecommand{\doi}{DOI:\discretionary{}{}{}\begingroup
  \urlstyle{rm}\Url}\fi

\bibitem[{Askanius(2021)}]{Askanius_2021}
Askanius T (2021) Memes and media’s role in radicalization.
\newblock \emph{The Journal of Intelligence, Conflict, and Warfare} 4(2):
  115–121.
\newblock \doi{10.21810/jicw.v4i2.3753}.

\bibitem[{Barakat and Rankin(2022)}]{Barakat_Rankin_2022}
Barakat M and Rankin S (2022) Youngkin looks to root out critical race theory
  in virginia.
\newblock
  \urlprefix\url{https://apnews.com/article/education-richmond-race-and-ethnicity-racial-injustice-virginia-8ad5da65b9cb05265f2b8081c41827cd}.

\bibitem[{Beauchamp(2021)}]{Beauchamp_2021b}
Beauchamp Z (2021) Did critical race theory really swing the virginia election?
\newblock
  \urlprefix\url{https://www.vox.com/policy-and-politics/2021/11/4/22761168/virginia-governor-glenn-youngkin-critical-race-theory}.

\bibitem[{Bonilla-Silva(1997)}]{Bonilla-Silva_1997}
Bonilla-Silva E (1997) Rethinking racism: Toward a structural interpretation.
\newblock \emph{American Sociological Review} 62(3): 465–480.
\newblock \doi{10.2307/2657316}.

\bibitem[{Burroughs(2020)}]{Burroughs_2020}
Burroughs B (2020) \emph{Fake memetics: Political rhetoric and circulation in
  political campaigns}.
\newblock The MIT Press.

\bibitem[{Carney(2016)}]{Carney_2016}
Carney N (2016) All lives matter, but so does race.
\newblock \emph{Humanity \& Society} 40(2): 180–199.
\newblock \doi{10.1177/0160597616643868}.

\bibitem[{Cassam(2021)}]{Cassam_2021}
Cassam Q (2021) \emph{Bullshit, Post-Truth, and Propaganda}.
\newblock Oxford University Press.

\bibitem[{Cestone et~al.(2022)Cestone, Jones, Harris, Quezada and
  Roest-Gyimah}]{Cestone_Jones_Harris_Quezada_Roest-Gyimah_2022}
Cestone LM, Jones LV, Harris M, Quezada N and Roest-Gyimah N (2022) Black
  americans’ social emotional responses to race-related discriminatory
  content on social media.
\newblock \emph{Journal of Ethnic \& Cultural Diversity in Social Work} :
  1–12\doi{10.1080/15313204.2022.2137716}.

\bibitem[{Cox(2022)}]{Cox_2022}
Cox K (2022) 10 facts about black republicans.
\newblock
  \urlprefix\url{https://www.pewresearch.org/short-reads/2022/11/07/10-facts-about-black-republicans/}.

\bibitem[{CrowdTangle(2022)}]{crowdtangle}
CrowdTangle T (2022) Crowdtangle. facebook, menlo park, california, {US}. list
  id: 1733230.
\newblock \emph{https://crowdtangle.com/} .

\bibitem[{Dean(2018)}]{Dean_2018}
Dean J (2018) Sorted for memes and gifs: Visual media and everyday digital
  politics.
\newblock \emph{Political Studies Review} 17(3): 255–266.
\newblock \doi{10.1177/1478929918807483}.

\bibitem[{Delgado and Stefancic(2023)}]{Delgado_Stefancic_2023}
Delgado R and Stefancic J (2023) \emph{Critical race theory an introduction}.
\newblock New York University Press.

\bibitem[{Edenberg and Hannon(2021)}]{Edenberg_Hannon_2021}
Edenberg E and Hannon M (2021) \emph{Political epistemology}.
\newblock Oxford University Press.

\bibitem[{Entman and Rojecki(2007)}]{Entman_Rojecki_2007}
Entman RM and Rojecki A (2007) \emph{The black image in the white mind: Media
  and race in America}.
\newblock University of Chicago Press.

\bibitem[{Ester et~al.(1996)Ester, Kriegel, Sander and
  Xu}]{10.5555/3001460.3001507}
Ester M, Kriegel HP, Sander J and Xu X (1996) A density-based algorithm for
  discovering clusters in large spatial databases with noise.
\newblock In: \emph{Proceedings of the Second International Conference on
  Knowledge Discovery and Data Mining}, KDD'96. AAAI Press, p. 226–231.

\bibitem[{Fairclough(2018)}]{Fairclough_2018}
Fairclough N (2018) \emph{Critical discourse analysis: The critical study of
  language}.
\newblock Routledge.

\bibitem[{Farid(2021)}]{farid2021overview}
Farid H (2021) An overview of perceptual hashing.
\newblock \emph{Journal of Online Trust and Safety} 1(1).

\bibitem[{Ferguson(2022)}]{Ferguson_2022}
Ferguson A (2022) Redefining antiracism: Learning from activists to sharpen
  academic language.
\newblock \emph{Sociology Compass} 17(1).
\newblock \doi{10.1111/soc4.13057}.

\bibitem[{Gantt~Shafer(2017)}]{GanttShafer_2017}
Gantt~Shafer J (2017) Donald trump’s “political incorrectness”:
  Neoliberalism as frontstage racism on social media.
\newblock \emph{Social Media + Society} 3(3): 205630511773322.
\newblock \doi{10.1177/2056305117733226}.

\bibitem[{Lankshear and Knobel(2019)}]{Lankshear_Knobel_2019}
Lankshear C and Knobel M (2019) Memes, macros, meaning, and menace: Some trends
  in internet memes.
\newblock \emph{The Journal of Communication and Media Studies} 4(4): 43–57.
\newblock \doi{10.18848/2470-9247/cgp/v04i04/43-57}.

\bibitem[{Lee-Won et~al.(2017)Lee-Won, White and
  Potocki}]{Lee-Won_White_Potocki_2017}
Lee-Won RJ, White TN and Potocki B (2017) The black catalyst to tweet: The role
  of discrimination experience, group identification, and racial agency in
  black americans’ instrumental use of twitter.
\newblock \emph{Information, Communication \& Society} 21(8): 1097–1115.
\newblock \doi{10.1080/1369118x.2017.1301516}.

\bibitem[{Lewis(2005)}]{Lewis_2005}
Lewis AK (2005) Black conservatism in america.
\newblock \emph{Journal of African American Studies} 8(4): 3–13.
\newblock \doi{10.1007/s12111-005-1000-1}.

\bibitem[{Low(2019)}]{Low_2019}
Low J (2019) A pragmatic definition of the concept of theoretical saturation.
\newblock \emph{Sociological Focus} 52(2): 131–139.
\newblock \doi{10.1080/00380237.2018.1544514}.

\bibitem[{Matamoros-Fernández(2017)}]{Matamoros-Fernández_2017}
Matamoros-Fernández A (2017) Platformed racism: The mediation and circulation
  of an australian race-based controversy on twitter, facebook and youtube.
\newblock \emph{Information, Communication \& Society} 20(6): 930–946.
\newblock \doi{10.1080/1369118x.2017.1293130}.

\bibitem[{McIntyre(2018)}]{McIntyre_2018}
McIntyre L (2018) \emph{Post-truth}.
\newblock MIT Press.

\bibitem[{Mendelberg(2001)}]{Mendelberg_2001}
Mendelberg T (2001) \emph{The race card: Campaign strategy, implicit messages,
  and the norm of Equality}.
\newblock Princeton University Press.

\bibitem[{Monga and Evans(2006)}]{1709989}
Monga V and Evans B (2006) Perceptual image hashing via feature points:
  Performance evaluation and tradeoffs.
\newblock \emph{IEEE Transactions on Image Processing} 15(11): 3452--3465.
\newblock \doi{10.1109/TIP.2006.881948}.

\bibitem[{Moody-Ramirez et~al.(2021)Moody-Ramirez, Tait and
  Bland}]{Moody-Ramirez_Tait_Bland_2021}
Moody-Ramirez M, Tait GB and Bland D (2021) An analysis of {G}eorge
  {F}loyd-themed memes.
\newblock \emph{The Journal of Social Media in Society} 10(2): 373–401.

\bibitem[{Morrow et~al.(2022)Morrow, Swire‐Thompson, Polny, Kopec and
  Wihbey}]{Morrow_Swire‐Thompson_Polny_Kopec_Wihbey_2022}
Morrow G, Swire‐Thompson B, Polny JM, Kopec M and Wihbey JP (2022) The
  emerging science of content labeling: Contextualizing social media content
  moderation.
\newblock \emph{Journal of the Association for Information Science and
  Technology} 73(10): 1365–1386.
\newblock \doi{10.1002/asi.24637}.

\bibitem[{Nieubuurt(2021)}]{Nieubuurt_2021}
Nieubuurt JT (2021) Internet memes: Leaflet propaganda of the digital age.
\newblock \emph{Frontiers in Communication} 5.
\newblock \doi{10.3389/fcomm.2020.547065}.

\bibitem[{Noble(2018)}]{Noble_2018}
Noble SU (2018) \emph{Algorithms of oppression: How search engines reinforce
  racism}.
\newblock New York University Press.

\bibitem[{Ray et~al.(2017)Ray, Brown, Fraistat and
  Summers}]{Ray_Brown_Fraistat_Summers_2017}
Ray R, Brown M, Fraistat N and Summers E (2017) Ferguson and the death of
  michael brown on twitter: \#blacklivesmatter, \#tcot, and the evolution of
  collective identities.
\newblock \emph{Ethnic and Racial Studies} 40(11): 1797–1813.
\newblock \doi{10.1080/01419870.2017.1335422}.

\bibitem[{Rietdijk and Archer(2021)}]{Rietdijk2021-RIEPFB}
Rietdijk N and Archer A (2021) Post-truth, false balance and virtuous
  gatekeeping.
\newblock In: Snow N and Vaccarezza MS (eds.) \emph{Virtues, Democracy, and
  Online Media: Ethical and Epistemic Issues}. Routledge.

\bibitem[{Rose(2017)}]{Rose_2017}
Rose J (2017) Brexit, trump, and post-truth politics.
\newblock \emph{Public Integrity} 19(6): 555–558.
\newblock \doi{10.1080/10999922.2017.1285540}.

\bibitem[{Ross and Rivers(2018)}]{Ross_Rivers_2018}
Ross AS and Rivers DJ (2018) Internet memes as polyvocal political
  participation.
\newblock In: Schill DJ and Hendricks JA (eds.) \emph{The presidency and social
  media: Discourse, disruption, and digital democracy in the 2016 presidential
  election}. Routledge.

\bibitem[{Schaffner et~al.(2018)Schaffner, Macwilliams and
  Nteta}]{Schaffner_Macwilliams_Nteta_2018}
Schaffner BF, Macwilliams M and Nteta T (2018) Understanding white polarization
  in the 2016 vote for president: The sobering role of racism and sexism.
\newblock \emph{Political Science Quarterly} 133(1): 9–34.
\newblock \doi{10.1002/polq.12737}.

\bibitem[{Shifman(2013)}]{Shifman_2013}
Shifman L (2013) Memes in a digital world: Reconciling with a conceptual
  troublemaker.
\newblock \emph{Journal of Computer-Mediated Communication} 18(3): 362–377.
\newblock \doi{10.1111/jcc4.12013}.

\bibitem[{West(1996)}]{West_1996}
West C (1996) Foreward.
\newblock In: Crenshaw K, Gotanda N and Peller G (eds.) \emph{Critical Race
  Theory: The Key Writings That Formed the Movement}. The New Press.

\bibitem[{Woods and Hahner(2020)}]{Woods_Hahner_2020}
Woods HS and Hahner LA (2020) \emph{Make America meme again the rhetoric of the
  alt-right}.
\newblock Peter Lang Publishing Inc. New York.

\bibitem[{Zannettou et~al.(2018)Zannettou, Caulfield, Blackburn, Cristofaro,
  Sirivianos, Stringhini and Suarez-Tangil}]{zannettou2018origins}
Zannettou S, Caulfield T, Blackburn J, Cristofaro ED, Sirivianos M, Stringhini
  G and Suarez-Tangil G (2018) On the origins of memes by means of fringe web
  communities.

\bibitem[{Zheng(2021)}]{Zheng_2021}
Zheng L (2021) It’s not your coworkers’ job to teach you about social
  issues.
\newblock
  \urlprefix\url{https://hbr.org/2019/07/its-not-your-coworkers-job-to-teach-you-about-social-issues}.

\end{thebibliography}

\end{document}